\documentstyle[seceq,epsf]{ptptex}

\def\PTPS#1{Prog.\ Theor.\ Phys.\ Suppl. \andvol{#1}}
\newcommand{\nc}{\newcommand}		
\nc{\nuc}[2]	{$^{#1}${#2}} 		
\nc{\vc}[1]	{\mbox{\boldmath $#1$}}	
\nc{\al}	{\alpha}		
\nc{\th}	{\theta}		
\nc{\lam}	{\lambda}		
\nc{\Gam}	{\Gamma}		
\nc{\gam}	{\gamma}		
\nc{\bra}	{\langle}		
\nc{\ket}	{\rangle}		
\nc{\bras}[1]	{\langle #1|}		
\nc{\kets}[1]	{|#1\rangle}		
\nc{\hO}	{\hat{O}}		
\nc{\wtil}	{\widetilde}		
\nc{\eps}	{\epsilon}		%
\nc{\mapleft}[1]{			
 \smash{\mathop{\,			%
  \hbox to 1.2cm{\rightarrowfill}\, }\limits_{#1}}}
\begin{document}
\markboth{
	T.~Myo, 
	S.~Aoyama,
	K.~Kat\=o,
	K.~Ikeda
}{
Extended \nuc{9}{Li}+$n$+$n$ three-body model of \nuc{11}{Li} with the pairing correlation in $^9$Li 
}
\title{
Extended \nuc{9}{Li}+$\mib{n}$+$\mib{n}$ three-body model of \nuc{11}{Li}\\with the pairing correlation in $^9$Li 
}
\author{Takayuki {\sc Myo},
	Shigeyoshi {\sc Aoyama}$^1$,
	Kiyoshi {\sc Kat\=o}
	and Kiyomi {\sc Ikeda}$^2$
}
\inst{	Division of Physics, Graduate School of Science,
	Hokkaido University, Sapporo 060-0810,\\
	$^1$Information Processing Center, Kitami Institute of Technology,
	Kitami 090-8507,\\
	$^2$RI-Beam Science Laboratory, RIKEN(The Institute of Physical and Chemical Research), 
	Wako, Saitama 351-0198, Japan.
}
\recdate{
	\today
}
\abst{
We discuss the binding mechanism of \nuc{11}{Li} 
based on an extended three-body model of \nuc{9}{Li}+$n$+$n$.
In the model, we take into account the pairing correlation of $p$-shell neutrons in \nuc{9}{Li}, 
in addition to that of valence neutrons outside the \nuc{9}{Li} nucleus,
and solve the coupled-channel two- and three-body problems of \nuc{10}{Li} and \nuc{11}{Li}, respectively.
The results show that degrees of freedom of the pairing correlation in \nuc{9}{Li} 
play an important role in the structure of \nuc{10}{Li} and \nuc{11}{Li}. 
In \nuc{10}{Li}, the pairing correlation in \nuc{9}{Li} produces a so-called pairing-blocking effect
due to the presence of valence neutron, which degenerates $s$- and $p$-wave neutron orbits energetically.
In \nuc{11}{Li}, on the other hand, the pairing-blocking effect is surpassed by the core-$n$ interaction
due to two degrees of freedom of two valence neutrons surrounding \nuc{9}{Li}, and as a result, 
the ground state is dominated by the $p$-shell closed configuration and does not show a spatial extension 
with a large r.m.s. radius.
These results indicate that the pairing correlation is realized differently 
in odd- and even-neutron systems of \nuc{10}{Li} and \nuc{11}{Li}.
We further improve the tail part of the \nuc{9}{Li}-$n$ interaction, which 
works well to reproduce the observed large r.m.s. radius in \nuc{11}{Li}.
}
\maketitle
\section{Introduction}\label{sec:intro}
Developments of radioactive beams provide us with many interesting phenomena
of unstable nuclei near the drip lines.\cite{Ta85,Ta88a,Ta96}
The most typical example is the discovery of a neutron halo structure observed 
in several neutron-rich nuclei such as \nuc{6}{He},~\nuc{11}{Li} and \nuc{11}{Be}.\cite{Ta85,Ta88a}
One of the common features of unstable nuclei is the weak binding; 
the neutron halo nuclei have extremely small binding energies against one- or two-neutron emission. 
This property of halo nuclei indicates 
a local breaking of saturations of densities and binding-energies observed in stable nuclei.
In unstable nuclei, most of excited states are unbound as a result of the weak binding.
It is expected that weakly-bound halo states 
have a strong influence on the properties of unbound states. 
The soft-dipole resonance\cite{Ha87,Ike92} is 
one of the most interesting problems concerning with 
a characteristic excitation mode arising from the weak-binding energy of neutron halo nuclei.

The \nuc{11}{Li} nucleus is well known as a typical two-neutron halo system 
with the small two-neutron separation energy; 0.31 MeV\cite{Au93} 
and the large matter radius of its ground state.
The understanding of the structure of \nuc{11}{Li} is very important 
to get the fundamental knowledge of neutron-rich nuclei.
So far, there are many studies on the low-energy structure of \nuc{11}{Li}. 
In addition to the halo structure of the ground state,
the excitation mechanism of \nuc{11}{Li} is also interesting 
to learn the electromagnetic properties such as the soft-dipole resonance.
One of the keys to understand such low-energy structures of \nuc{11}{Li} 
is $1s$-wave component mixing of valence neutrons.
The large matter radius of the \nuc{11}{Li} ground state implies 
a large mixing of the $(1s_{1/2})^2$-components in addition to the $(0p_{1/2})^2$ ones in the wave function
(This mixing is also important to discuss the possible excited states in \nuc{11}{Li}).
Analysis of Gamow-Teller transition\cite{Su94} and a recent fragmentation experiment\cite{Si99} 
of \nuc{11}{Li} also suggest the same trend.
Observations of excited states in \nuc{11}{Li}, which may be related to the $s$-wave, 
were reported experimentally,\cite{Ko92,Bo95,Ko96,Ko97,Go98}
and also discussed in the theoretical studies,\cite{Ka97,Su00,Ao02}
although the conclusive results are not obtained yet.
In \nuc{12}{Be} having the same neutron number as \nuc{11}{Li}, 
the similar discussion of the existence of the $s$-orbit in the low excitation energy region 
has been done for the ground state,\cite{Na00,Iw00a,It00}
and for a low-lying $1^-$ excited state.\cite{Iw00b,Sa01}
This low-lying $s$-wave state breaks the magic number $N=8$, the $p$-shell
closed configuration, in neutron-rich nuclei.

Related to these facts found in the nuclei having even neutron number,
the $s$-wave states are also discussed to appear in a low-energy structure of \nuc{10}{Li}, 
having odd number of neutrons. 
So far, there are many experimental studies on the spectroscopy of 
\nuc{10}{Li}.\cite{Wi75,Am90,Kr93,Bo93,Yo94,Zi95,Zi97,Bo99,Th99,Cha01,Che01}
Several experiments suggest $s$-wave states ($1^-$ and/or $2^-$) near 
the threshold energy of \nuc{9}{Li}+$n$,\cite{Am90,Zi95,Th99,Cha01} 
and it implies the degeneracy of $s$- and $p$-waves in \nuc{10}{Li}.
Furthermore, in \nuc{9}{He}, having the same neutron number as that of \nuc{10}{Li}, 
recently, there is an experimental suggestion of the $s$-wave ground state.\cite{Che01}
Systematically, these phenomena of low energy $s$-waves seen in \nuc{9}{He}, \nuc{10}{Li} and 
\nuc{11}{Be} are related to the inversion problem of $N=7$ isotone\cite{Th99,Sa93} 
as will be discussed later again.
The essential mechanism to lower the $s$-wave around the threshold energy 
in these nuclei is still unclear, although degrees of freedom of deformation and 
coupling to the core's excited states have been discussed for \nuc{11}{Be}.

So far, there are many theoretical studies of \nuc{11}{Li} 
based on the three-body model of \nuc{9}{Li}+$n$+$n$.\cite{Ba77,To90a,Su90a,To90b,Be91,Es92,Ba92,To94,De97,Ba97,Mu98}
In some of them, the state-dependent \nuc{9}{Li}-$n$ interaction are often adopted. 
Thompson and Zhukov\cite{To94} first proposed such an interaction
where the $s$-wave state comes down around the threshold energy enough to be a virtual state in \nuc{10}{Li}.  
Without any theoretical explanation, they used the common \nuc{9}{Li}-$n$ interaction 
in the calculation of \nuc{11}{Li} as that of \nuc{10}{Li}, 
and discussed the importance of the virtual $s$-wave states in \nuc{10}{Li} 
on the halo structure of \nuc{11}{Li} with a large mixing of the $(1s_{1/2})^2$-component. 
Calculations of the breakup reactions using such a state-dependent interaction,
have been reported.\cite{To94,Zu93,Co97,Ga97,Ga99,Ga01a,Ga01b}
Although the properties of \nuc{11}{Li} and \nuc{10}{Li} are consistent with an experimental situation,
the mechanism to explain the state-dependency in the \nuc{9}{Li}-$n$ interaction
and the reliability to use the common interaction between two nuclei are not realized yet.
The consistent understanding of the binding mechanism of \nuc{11}{Li} 
and the spectroscopy of \nuc{10}{Li} is still remaining as a basic question.
It is, therefore, necessary to discuss the availabilities of the state-dependency 
and the common uses of the \nuc{9}{Li}-$n$ interaction for the analysis of \nuc{10}{Li} and \nuc{11}{Li}.
 
In our first studies of \nuc{10}{Li}\cite{Ka93} and \nuc{11}{Li},\cite{Mu98} 
we employed the simple \nuc{9}{Li}+$n$+$n$ model with adopting the frozen \nuc{9}{Li} core,
where we used the state-independent \nuc{9}{Li}-$n$ interaction to reproduce the experimental 
$p$-wave resonances but not to have the low-energy $s$-state in \nuc{10}{Li}.
Results show that the binding energy of the \nuc{11}{Li} ground state is short of about 1 MeV 
(0.7 MeV measured from the three-body threshold energy) in comparison to experimental one.
We consider that this underbinding problem is related to the assumption of the three-body model.
In Ref.~\citen{Mu98}, we assumed the single shell model 
configuration of the $0p_{3/2}$ sub-closed neutrons in \nuc{9}{Li}.
Indeed such a description of \nuc{9}{Li} is acceptable as a assumption,
however, not strongly allowable rather than the case of $(0s_{1/2})^4$ configuration 
of \nuc{4}{He} in the three-body analysis of \nuc{6}{He}.

One of the ways to improve the three-body model of \nuc{9}{Li}+$n$+$n$
is to take a configuration mixing of \nuc{9}{Li} considering the pairing correlation of $p$-shell neutrons. 
This indicates that we can regard the $p$-shell neutrons in \nuc{9}{Li} as a member of valence neutrons 
in consideration of the Pauli principle between valence neutrons inside and outside the \nuc{9}{Li} nucleus.
In this sense, the \nuc{9}{Li} nucleus is treated as an active core 
including the internal degrees of freedom of the neutron pairing correlation.
We also call the last two valence neutrons outside the \nuc{9}{Li} nucleus, 
as active valence neutrons to distinguish from the valence neutrons in \nuc{9}{Li},
because their wave function is dynamically solved 
in a few-body approach as will be explained in the next section.
We already learned that the pairing correlation between the active valence neutrons 
plays an important role in the weak binding of \nuc{11}{Li}\cite{Mu98} and \nuc{6}{He}.\cite{Ao95}
It is, therefore, natural to consider that the pairing correlation between valence neutrons 
in the \nuc{9}{Li} core nucleus might be also important in the binding mechanism of neutron halo nuclei,
through the coupling with the active valence neutron pair.

\begin{figure}[t]
\begin{center}
\epsfxsize=11.5cm
\centerline{\epsfbox{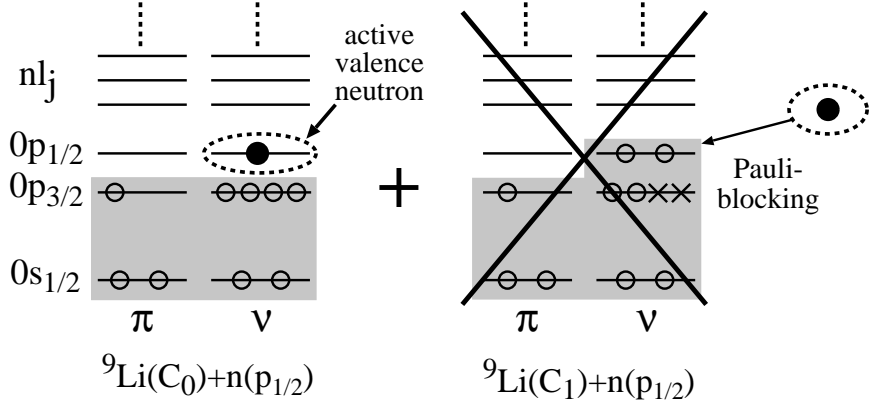}}
\caption[Schematic illustration to explain the pairing-blocking]{
Schematic illustration to explain the pairing-blocking effect in the \nuc{9}{Li}-$n$($p$-wave) system, 
where $C_0$ and $C_1$ correspond to the neutron sub-closed configuration and pairing excited one, respectively.}
\label{fig:Pauli}
\end{center}
\end{figure}

In \nuc{10}{Li}, the pairing correlation in \nuc{9}{Li} connects to the 
so-called pairing-blocking due to the presence of the $p$-wave active valence neutron
shown in Fig.~\ref{fig:Pauli}.
In a core-plus-valence neutron system, 
the pairing excited configurations in the core nucleus suffer the blocking effect 
from the active valence neutron due to the Pauli-principle associated with the occupied orbits.
Sagawa et al.\cite{Sa93} also pointed the importance of the pairing-blocking effect 
in order to explain the level inversion of $1/2^+$ and $1/2^-$ states in \nuc{11}{Be}
in addition to the degrees of freedom of the core excitation.

In our second study of \nuc{10}{Li},\cite{Ka99} 
we used this idea to explain the low-energy $s$-waves in \nuc{10}{Li} with the \nuc{9}{Li}+$n$ model.
We performed the configuration mixing of $C_1~:~(0p_{3/2})^2_\nu(0p_{1/2})^2_\nu$
in addition to $C_0~:~(0p_{3/2})^4_\nu$ in the ground state of the \nuc{9}{Li} cluster
and solved the coupled-channel two-body problem of \nuc{9}{Li}+$n$ explained in Fig.~\ref{fig:Pauli}.
Due to the pairing-blocking, the $p$-wave state of \nuc{10}{Li} is pushed up energetically,
and as a result, the energy distance between $s$- and $p$-waves becomes small.
The decay widths of the $p$-wave resonances ($1^+,2^+$) are significantly improved
in the coupled-channel calculation,  
and also the $s$-wave states come down around the threshold of \nuc{9}{Li}+$n$ 
to be the virtual states.\cite{Ma00} 
These results are consistent with the recent experimental situation,
and indicate that the pairing correlation in the \nuc{9}{Li} core
plays an essential role in understanding the structure of \nuc{10}{Li}.
In Ref.~\citen{Ka99}, we also derived the effective state-dependent \nuc{9}{Li}-$n$ interactions
for $s$- and $p$-waves by renormalizing the pairing-blocking effect into the interaction,
in which the $s$-wave interaction is deeper than that of the $p$-wave.
We showed that the pairing-blocking effect can be a dynamical origin of 
the state dependency in the \nuc{9}{Li}-$n$ interaction.

In Ref.~\citen{Ao02}, we carried out the three-body calculation of \nuc{11}{Li}
using an effective state-dependent \nuc{9}{Li}-$n$ interaction
reflecting the pairing-blocking effect, which was constructed in Ref.~\citen{Ka99}. 
The calculated results well reproduced the ground state properties of $^{11}$Li.\cite{Ao02}
However, when the repulsive potential for the $p$-wave is assumed to be half in $^{11}$Li from that in $^{10}$Li,
in order to avoid a double count of the pairing-blocking in \nuc{9}{Li}+$n$+$n$, 
we obtain the $p$-shell dominant ground state of \nuc{11}{Li}.
It is required to do a microscopic treatment of the pairing
correlation of \nuc{9}{Li} in the three-body calculation of \nuc{11}{Li},
for the purpose to make clear the ground of these effective treatment of 
the \nuc{9}{Li}-$n$ interactions in \nuc{10}{Li} and \nuc{11}{Li}.

In this study, we proceed our study to the extended \nuc{9}{Li}+$n$+$n$ three-body analysis of \nuc{11}{Li} 
by taking into account the pairing correlation in \nuc{9}{Li}.
We see how the pairing correlation of valence neutrons affects the binding mechanism of \nuc{11}{Li}.
It is very interesting to see whether this approach is satisfactory
for the simultaneous understanding of \nuc{10}{Li} and \nuc{11}{Li}.
We verify the state-dependency of the \nuc{9}{Li}-$n$ interaction
arising from the pairing correlation of the \nuc{9}{Li} core
in the three-body calculation of \nuc{11}{Li}, as was done in the case of \nuc{10}{Li}.
We also discuss the couplings between the pairing correlation
of valence neutrons in \nuc{9}{Li} and that of active valence neutrons,
and the difference between their roles in odd- and even-neutron systems 
of \nuc{10}{Li} and \nuc{11}{Li}.
The merit of the three-body model is that 
we can obtain the three-body eigenstates and two-body ones of subsystems
accurately including resonances with proper boundary conditions for the particle emission
in the framework of the complex scaling method.\cite{ABC}
In this paper, we focus on the structure of the \nuc{11}{Li} ground state, 
such as the binding energy and the probability of the $(1s_{1/2})^2$-component.

In \S~\ref{sec:method}, we explain how to treat the pairing correlation in \nuc{9}{Li}
in the extended three-body model of \nuc{11}{Li}. 
In \S~\ref{sec:result}, we show the results of \nuc{10}{Li} and \nuc{11}{Li} and  
discuss the role of the pairing correlation in the structures of these two nuclei.
In \S~\ref{sec:discuss}, we further discuss the tail effect in the \nuc{9}{Li}-$n$ interaction.
A Summary is given in \S~\ref{sec:summary}.

\section{Coupled-channel model of core+active valence neutrons system}\label{sec:method} 

We write \nuc{11}{Li} with a coupled-channel \nuc{9}{Li}+$n$+$n$ three-body model,
in which we adopt the multi-configuration representation for the \nuc{9}{Li} core
in order to take into account the pairing correlation in \nuc{9}{Li}.
This is a natural extension from the simple three-body model in Ref.~\citen{Mu98}, 
where a single shell model configuration for \nuc{9}{Li} is assumed.
Our previous results show that
single configuration of \nuc{9}{Li} is not good enough to reproduce the various properties of \nuc{11}{Li}.
We improve the description of the \nuc{9}{Li} core as was mentioned in \S~\ref{sec:intro}.

In the case of \nuc{6}{He} which is considered to have a simpler structure than that of \nuc{11}{Li},
based on a \nuc{4}{He}+$n$+$n$ model, 
dissociation or excitation of the \nuc{4}{He} core in \nuc{6}{He} has been discussed 
in some studies.\cite{Cs93,Ar99}
The contribution from the dissociation of \nuc{4}{He} 
is expected to be very small because the \nuc{4}{He} core is rigid.
The contribution to the binding energy of \nuc{6}{He} is about 0.2 MeV 
which is one-order smaller than the energy of the relative motion between core and valence neutrons.
It has been discussed to treat the contribution from the \nuc{4}{He}-dissociation 
in the \nuc{4}{He}+$n$+$n$ system 
by introducing the effective three-body \nuc{4}{He}-$n$-$n$ interaction phenomenologically.\cite{My01}
Such a perturbative treatment was shown to reproduce successfully the properties of \nuc{6}{He} 
such as energy spectra and matter radius.

On the contrary, since \nuc{9}{Li} is not considered to be a rigid core as \nuc{4}{He}, 
because one-neutron separation energy in \nuc{9}{Li} is about 4 MeV,
we must discard the assumption of the single-closed shell configuration 
for neutrons in the \nuc{9}{Li} cluster and take into account the multi-configuration mixing.
In order to do this, we pay attention to the pairing correlation in the $p$-shell neutrons. 
As was already mentioned, the pairing-blocking effect coming from the pairing correlation in \nuc{9}{Li} 
explains the degeneracy of $s$- and $p$-waves in the \nuc{9}{Li}+$n$ system.
It is, therefore, very interesting to investigate \nuc{11}{Li} with the pairing correlation in \nuc{9}{Li}
as we did in the analysis of \nuc{10}{Li}.
In this section, we briefly explain how to introduce the pairing correlation in \nuc{9}{Li},
and apply it to the three-body model of \nuc{11}{Li}. 
Detailed explanations for the two-body system \nuc{10}{Li}, are given in Ref.~\citen{Ka99}.

\subsection{Neutron pairing correlation in \nuc{9}{Li}}\label{sec:9Li}
The wave function of the \nuc{9}{Li} ground state ($J^\pi=3/2^-$) is assumed 
by the following linear combination of pairing configuration:
\begin{eqnarray}
	\Phi(^{9}{\rm Li})
&=&	\sum_{\al=0}^{N_\al}\ a_\al\ \Phi(C_\al),
	\label{eq:WF_9Li}
	\\
	\Phi(C_0)
&:&	(0s_{1/2})^4 (0p_{3/2})_\pi (0p_{3/2})^4_\nu
	\nonumber
	\\
	\Phi(C_1)
&:&	(0s_{1/2})^4 (0p_{3/2})_\pi (0p_{3/2})^2_\nu (0p_{1/2})^2_\nu 
	\nonumber
	\\
	\Phi(C_2)
&:&	(0s_{1/2})^4 (0p_{3/2})_\pi (0p_{3/2})^2_\nu (1s_{1/2})^2_\nu 
	\nonumber
	\\
	\vdots\hspace*{0.3cm}
&&	\hspace*{2.0cm}\vdots
	\nonumber
	\\
	\Phi(C_\al)
&:&	(0s_{1/2})^4 (0p_{3/2})_\pi (0p_{3/2})^2_\nu (nl_j)^2_\nu .
	\nonumber
\end{eqnarray}
Here, $\al$ and $N_\al$ denotes the label to distinguish each configuration of \nuc{9}{Li}, 
and the number of considering configurations, respectively. 
We assume only the $j^\pi=0^+$ pairing configuration of core neutrons
as $[(0p_{3/2})(0p_{3/2})]_{0^+}[(nl_j)(nl_j)]_{0^+}$.
The amplitude $a_\al$ is determined by solving the Schr\"odinger equation 
for an isolated \nuc{9}{Li} core given as:
\begin{eqnarray}
	H(^{9}{\rm Li})\Phi(^{9}{\rm Li})
&=&	E(^{9}{\rm Li})\Phi(^{9}{\rm Li}).
	\label{eq:H_9Li}
\end{eqnarray}
The Hamiltonian of \nuc{9}{Li} is given in a matrix form as
\begin{eqnarray}
	H(^{9}{\rm Li})
&=&	\left(
	\begin{array}{ccccc}
\wtil{G}_{0p_{3/2},0p_{3/2}} & \wtil{G}_{0p_{3/2},0p_{1/2}} & \wtil{G}_{0p_{3/2},1s_{1/2}} & \cdots & \wtil{G}_{0p_{3/2},nl_j} \\
\wtil{G}_{0p_{1/2},0p_{3/2}} & \wtil{G}_{0p_{1/2},0p_{1/2}} & \wtil{G}_{0p_{1/2},1s_{1/2}} & \cdots & \wtil{G}_{0p_{1/2},nl_j} \\
\wtil{G}_{1s_{1/2},0p_{3/2}} & \wtil{G}_{1s_{1/2},0p_{1/2}} & \wtil{G}_{1s_{1/2},1s_{1/2}} & \cdots & \wtil{G}_{1s_{1/2},nl_j} \\
     \vdots           &  	     \vdots   &   \vdots              &        &   \vdots        \\
\wtil{G}_{nl_j,0p_{3/2}} & \wtil{G}_{nl_j,0p_{1/2}} & \wtil{G}_{nl_j,1s_{1/2}} & \cdots & \wtil{G}_{nl_j,nl_j} \\
	\end{array}
	\right),
	\label{eq:ham_9Li}
	\nonumber
	\\
	\\
	\wtil{G}_{nl_j,n'l'_{j'}}
&=&	G_{nl_j,n'l'_{j'}}~+~2\ \Delta E_{nl_j}\cdot \delta_{n,n'} \delta_{l,l'} \delta_{j,j'} ,
	\label{eq:ham_9Li_2}
	\\
	G_{nl_j,n'l'_{j'}}
&=&	\bra (nl_j)^2_\nu|v^G_{nn}| (n'l'_{j'})^2_\nu\ket,\qquad
	\Delta E_{nl_j}
~=~	\eps_{nl_j}-\eps_{0p_{3/2}}.
\end{eqnarray}
Here, $v^G_{nn}$ is a neutron-neutron interaction to determine the pairing matrix element 
in \nuc{9}{Li}.
The $\eps_{nl_j}$ and $\Delta E_{nl_j}$ represent a position of single particle energies of 
$nl_j$-orbits in \nuc{9}{Li} 
and a difference of single particle energies between $0p_{3/2}$- and $nl_j$-orbits.
In the case of $nl_j=0p_{1/2}$, $\Delta E_{nl_j}$ represents the LS splitting in the $p$-shell.
The LS-splitting is determined in the same manner as in Ref. \citen{Ka93},
where we consider that the observed level spacing between $1/2^-$ and $3/2^-$ states of \nuc{11}{Be} 
expresses the $0p_{3/2}$-$0p_{1/2}$ splitting.
The size of \nuc{9}{Li} is set to reproduce the experimental matter radius 
$2.32\pm0.02$ fm of \nuc{9}{Li},\cite{Ta88b} which leads to the length parameter 
$b$=1.69 fm of the harmonic oscillator wave function.

In this paper, we restrict ourselves to consider the neutron pairing correlation in \nuc{9}{Li} 
up to $(0p_{1/2})^2$ component,
because we checked the contributions of other higher orbits (above the sd-shell) were very small.
For example, for the $1s$-orbit, the coupling matrix element $\wtil{G}_{0p_{3/2},1s_{1/2}}$ has a tendency 
to be very small with several $NN$ interactions, 
and its mixing is also very small in the \nuc{9}{Li} ground state.\cite{Ka99}
Other higher orbits are unbound states 
above the threshold of particle emission from \nuc{9}{Li}, 
and contributions from such unbound configurations are also small in the ground state of \nuc{9}{Li}.
Then we consider the configuration mixing of such higher orbits are less important 
in the following discussion except for the $0p_{1/2}$-orbit, 
which gives the essential contribution in the present analysis. 
It can be said that we include the small contributions of higher orbits 
into the matrix elements of the $0p_{1/2}$-orbit effectively. 

\begin{table}[t]
\caption[Pairing matrix elements in $0p$-shell.]{
Pairing matrix elements (in MeV) in the $0p$-shell and the LS-splitting using various $NN$ interactions; 
GPT(C): GPT interaction including only the central term, GPT(C+LS+T): GPT interaction including central, ls and tensor
terms.
}\label{tab:pairing-core}
\begin{center}
\begin{tabular}{c|c|c|c|c}
\hline
\hline
$NN$ interaction 
& $\wtil{G}_{0p_{3/2},0p_{3/2}}$ & $\wtil{G}_{0p_{3/2},0p_{1/2}}$ & $\wtil{G}_{0p_{1/2},0p_{1/2}}$ 
& $\Delta E_{0p_{1/2}}$\\
\noalign{\hrule height 0.5pt}
HN-1            &  $-$3.93    & $-$2.78	 &  1.28  & 1.62 \\
MHN             &  $-$3.95    & $-$2.76	 &  1.23  & 1.65 \\
GPT(C)          &  $-$3.59    & $-$2.57  &  1.76  & 1.77 \\
GPT(C+LS+T)     &  $-$2.81    & $-$3.67  &  1.62  & 0.92 \\
KYI\cite{Ka99}  &  ~0.0       & $-$5.62  &  6.46  & 3.23 \\
\hline
\end{tabular}
\end{center}
\vspace*{0.7cm}
\caption[Results of the amplitudes of the configuration mixing in \nuc{9}{Li}]{
Results of the mixing amplitudes $a_\al$ of each configuration of \nuc{9}{Li} in Eq.~(\ref{eq:WF_9Li})
and the energy gains with unit in MeV for energy.
$NN$ interactions are the same as Table~\ref{tab:pairing-core}.}
\label{tab:pairing-mix}
\begin{center}
\begin{tabular}{c|c|c|c|c}
\hline
\hline
$NN$ interaction  &  $(a_0)^2$  & $(a_1)^2$ & energy gain & label  \\
\noalign{\hrule height 0.5pt}
HN-1           &~~0.842~~&~~0.158~~& $-$1.20 & --- \\
MHN            &~~0.845~~&~~0.155~~& $-$1.18 & PC-W\\
GPT(C)         &~~0.860~~&~~0.140~~& $-$1.04 & --- \\
GPT(C+LS+T)    &~~0.759~~&~~0.241~~& $-$2.07 & --- \\
KYI\cite{Ka99} &~~0.750~~&~~0.250~~& $-$3.25 & PC-S\\
\hline
\end{tabular}
\end{center}
\end{table}

In order to calculate the pairing matrix element $v^G_{nn}$,
we use several kinds of the $NN$ interactions such as 
Hasegawa-Nagata No.1 (HN-1) and Modified Hasegawa Nagata (MHN)\cite{Mu98,Ka93,Ka99,Fu80}
and GPT\cite{Zu93,Co97,Go70} including only the central term and 
GPT including central, ls and tensor terms.
We also compare them with the result of Ref.~\citen{Ka99}, named ``KYI'' here, 
in which the Cohen-Krush interaction was used as a reference.
In the KYI parameter, the value of $\wtil{G}_{0p_{3/2},0p_{3/2}}$ is taken as the origin of energy.
In Table~\ref{tab:pairing-core}, we list the matrix elements of the $0p$-shell.
Estimated LS-splittings of the $p$-shell are 1.62 MeV, 1.65 MeV and 0.92 MeV for HN-1, MHN and GPT, 
respectively.

We list the configuration mixing in \nuc{9}{Li} in Table~\ref{tab:pairing-mix}, determined
by solving the eigenvalue problem of Eq.~(\ref{eq:H_9Li}).
It is found that HN-1 and MHN interactions give the similar values 
of the configuration mixing in \nuc{9}{Li} around 15\% for the pairing excited configuration.
The central part of GPT (fourth row) also gives the similar result to those of HN-1 and MHN cases.
On the other hand, the GPT interaction including tensor and spin-orbit terms (fifth row)
leads to a stronger mixing than the former three cases, and it is very similar to the KYI case.
The strong mixing is caused by the large coupling matrix element of $\wtil{G}_{0p_{3/2},0p_{1/2}}$, 
and we checked that the tensor force mainly contributes to produce a large coupling in the GPT interaction.
It can be said that the KYI parameter set effectively includes the contribution of 
tensor force in their matrix elements.
 
In the following analysis of \nuc{10}{Li} and \nuc{11}{Li}, 
we adopt the two kinds of mixing values determined using MHN and KYI for \nuc{9}{Li},
as typical cases to see configuration mixing effects due to the pairing correlation.
We call the KYI case as ``PC-S'' (strong case of pairing interaction) and 
the MHN case as ``PC-W'' (weak case of pairing interaction), hereafter.
This difference also gives a direct influence on the Pauli-blocking 
due to the presence of an active valence neutron in \nuc{10}{Li}.

\subsection{Extended three-body model of \nuc{11}{Li} with OCM} 
We solve the three-body problem of \nuc{9}{Li}+n+n with 
the orthogonality condition model (OCM).
The Hamiltonian of the present model is given as follows:
\begin{eqnarray}
	H(\mbox{\nuc{11}{Li}})
&=&	H(\mbox{\nuc{9}{Li}})
+	\sum_{i=1}^3{t_i} - T_G
+	\sum_{i=1}^2V_{cn}(\vc{r}_i) + V_{nn} 
+	\lambda\, \kets{\phi_{PF}}\bras{\phi_{PF}},\
	\label{eq:Ham}
\end{eqnarray}
where $H(\mbox{\nuc{9}{Li}})$, $t_i$ and $T_G$ are 
the internal Hamiltonian of \nuc{9}{Li} defined in Eq.~(\ref{eq:ham_9Li}),
kinetic energies of each clusters and the center-of-mass of the three-body system, 
respectively.
The two-body interaction between \nuc{9}{Li} and an active valence neutron, $V_{c n}$, is 
taken as a folding-type potential with the MHN interaction.
The Minnesota interaction\cite{Ta78} is used for two active valence neutrons
where the exchange mixture $u$ is chosen to be 0.95.
These choices are the same as Refs.~\citen{Mu98},~\citen{Ka99}~and~\citen{Ao97}.
The folding potential for \nuc{9}{Li}-$n$ includes the coupling between 
intrinsic spins of the active valence neutron and \nuc{9}{Li}($3/2$),
and this coupling produces the splittings of the energy levels, 
for instance $1^+$-$2^+$ (for $p_{1/2}$-neutron) and $1^-$-$2^-$ (for $s_{1/2}$-neutron) 
in the \nuc{10}{Li} spectra.
The last term $\lambda\,\kets{\phi_{PF}}\bras{\phi_{PF}}$ presents a projection operator 
to remove the Pauli forbidden (PF) states from the \nuc{9}{Li}-$n$ relative motion.\cite{Ku86}
In this model, PF states for the relative motion depend on the configuration 
$C_\al$ of \nuc{9}{Li}, namely the occupied orbits of $C_\al$ by neutrons in the \nuc{9}{Li} cluster:
The PF states are given as
\begin{eqnarray}
	\phi_{PF}
&=&	\left\{
	\begin{array}{ll}
	0s_{1/2},~0p_{3/2}           &~~\mbox{for}\quad C_0\\
	0s_{1/2},~0p_{3/2},~0p_{1/2} &~~\mbox{for}\quad C_1\\
	0s_{1/2},~0p_{3/2},~1s_{1/2} &~~\mbox{for}\quad C_2\\
	\hspace*{1.6cm}\vdots &~~~\vdots  \\
	0s_{1/2},~0p_{3/2},~nl_j     &~~\mbox{for}\quad C_{\al},\\
	\end{array}
	\right.
\end{eqnarray}
where the value of $\lambda$ is taken as $10^6$~MeV in this calculation.

The wave function of \nuc{11}{Li} is given as: 
\begin{eqnarray}
	\Psi^J(^{11}{\rm Li})
&=&	\sum_\al^{N_\al}
        {\cal A}\left\{\, [\Phi^{3/2^-}(C_\al), \chi^{j}_\al(nn)]^J\,\right\}.
	\label{eq:WF_11Li}
\end{eqnarray}
Here, $\chi^{j}_{\al}(nn)$ expresses the wave functions of two active valence neutrons,
and $j$ and $J$ are the spin of two active valence neutrons and the total spin of \nuc{11}{Li}, respectively.
The index $\al$ is the same as defined in Eq.~(\ref{eq:WF_9Li}).

The motion of the weak-binding active valence neutrons around the \nuc{9}{Li} core must be solved 
accurately on the basis of recent developments of few-body problem.
We employ here a variational method and the basis functions of the so-called 
hybrid-TV model,\cite{Ike92,To90b,Mu98,Ao95,My01}
where relative wave functions of \nuc{9}{Li}+$n$+$n$ are 
expanded with the combination of the basis states of the cluster orbital shell model 
(COSM; V-type),\cite{Su88,Su90}
and those of the extended cluster model (ECM; T-type),\cite{Ike92,To90b,Mu98} 
as follows:
\begin{eqnarray}
	\chi^j_\al(nn)
&=&	\chi^j_{\al,V}(\vc{\xi}_V)+\chi^{j}_{\al,T}(\vc{\xi}_T),
\end{eqnarray}
where $\vc{\xi}_V$ and $\vc{\xi}_T$ are V-type and T-type coordinate sets, respectively. 
In particular, ECM is important to take into account the pairing correlation 
between active valence neutrons,
which needs a very large COSM basis states to describe.\cite{Ike92,To90b,Mu98,Ao95}
The radial component of each relative wave function is expanded 
with a finite number of Gaussian centered at the origin, 
and the width parameters are chosen as a geometric progression.\cite{Ka88}
The three-body eigenstates are obtained by solving the eigenvalue problem of 
the coupled-channel Hamiltonian given in Eq.~(\ref{eq:Ham}).

Here, we briefly mention about the coupling between \nuc{9}{Li} configurations 
and the motion of active valence neutrons.
In \nuc{11}{Li}, the mixing probabilities of each configuration $C_\al$ of \nuc{9}{Li} are 
determined variationally to minimize the energy of the \nuc{11}{Li} ground state 
including the degrees of freedom of two active valence neutrons.
They depend on the relative coordinate between \nuc{9}{Li} and two active valence neutrons.
Asymptotically, when two active valence neutrons are far away from \nuc{9}{Li}, 
the wave function of \nuc{11}{Li} becomes as follows:
\begin{eqnarray}
	\chi^j_\al(nn)
&~~	\mapleft{r_1,r_2\to\infty}~~&\chi^j(nn),
	\\
	\Phi^J(^{11}{\rm Li})
&~~	\mapleft{r_1,r_2\to\infty}~~&
	\left[
	\left(\sum_\al^{N_\al} a_\al \Phi^{3/2^-}(C_\al) \right),~
	\chi^j(nn)
	\right]^J.
	\label{eq:asympt}
\end{eqnarray}
The first equation means that the asymptotic wave function of two active valence neutrons does not depend 
on the configuration of \nuc{9}{Li}, 
namely the coupling between active valence neutrons and \nuc{9}{Li} disappears
(The correlation between active valence neutrons also disappears).
The mixing amplitudes $\{a_\al\}$ of \nuc{9}{Li} in Eq.~(\ref{eq:asympt}) are
the same as that of the isolated \nuc{9}{Li} shown in Table~\ref{tab:pairing-mix}.
On the other hand, when two active valence neutrons are close to the \nuc{9}{Li} core, 
the motions of two active valence neutrons dynamically couple to the configuration of \nuc{9}{Li}
satisfying the Pauli principle,
which changes the mixing amplitudes $a_\al$ in \nuc{9}{Li} from those of the isolated \nuc{9}{Li} core.

\subsection{Exchange coupling between $2n$ in \nuc{9}{Li} and active valence $2n$}
\label{sec:exchange}

\begin{figure}[t]
\begin{center}
\epsfxsize=12cm
\centerline{\epsfbox{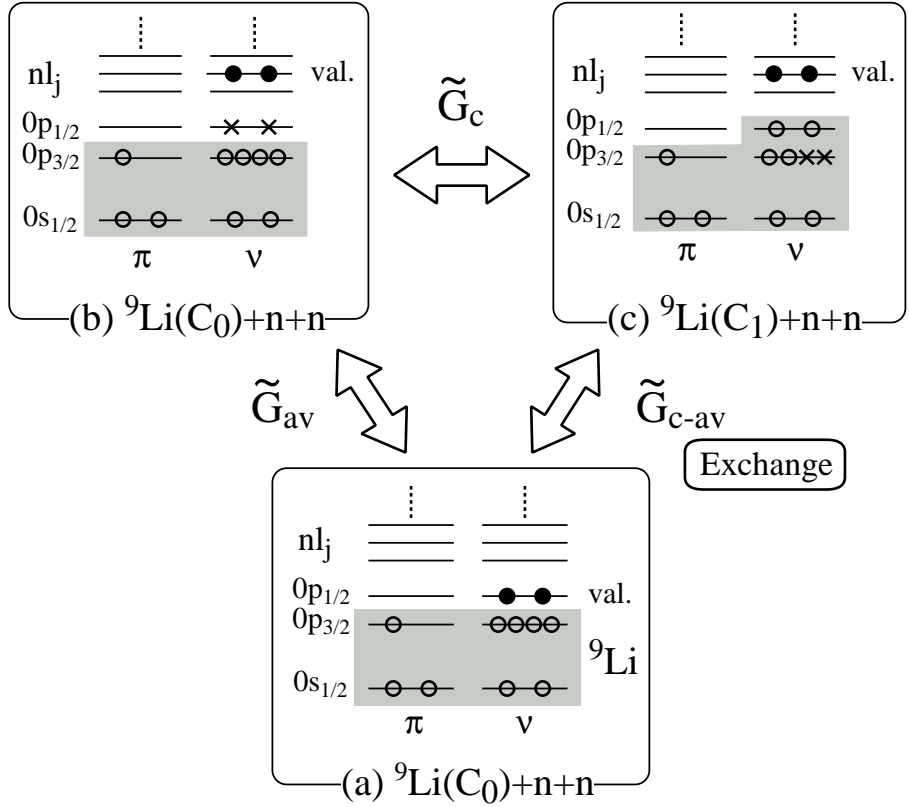}}
\caption[]{
Coupling schemes in \nuc{11}{Li} via the pairing interaction $\widetilde{G}$ 
between two neutrons in \nuc{9}{Li} and active valence neutrons.
Open circles are nucleons in the \nuc{9}{Li} core, and solid circles are active valence neutrons.}
\label{fig:coupling}
\end{center}
\end{figure}

The important coupling between \nuc{9}{Li} and active valence neutrons 
in the \nuc{11}{Li} ground state is illustrated in Fig.~\ref{fig:coupling}.
We categorize the configurations of \nuc{11}{Li} into three parts of (a), (b) and (c),
where a shaded area shows the part of \nuc{9}{Li}.
The panel (a) shows the lowest $p$-shell closed configuration, 
(b) the pairing configuration of two active valence neutrons excited from $(p_{1/2})^2$ to $(nl_j)^2$, 
and (c) a coupled configurations of two kinds of pairing excitations of two active valence neutrons 
and two valence neutrons in the \nuc{9}{Li} core.
In the extended three-body calculation of \nuc{11}{Li}, 
these three configurations are coupled through the pairing interaction
$\wtil{G}_{\rm pair}$ which can be separated into the following terms:
\begin{eqnarray}
	\wtil{G}_{\rm pair}
&=&	\wtil{G}_{\rm c}
+	\wtil{G}_{\rm av}
+	\wtil{G}_{\rm c-av}\, ,
\end{eqnarray}
where $\wtil{G}_{\rm c}$ is the same as $\wtil{G}_{nl_j,n'l'_{j'}}$ 
defined in Eq.~(\ref{eq:ham_9Li_2}) for the valence neutrons in the \nuc{9}{Li} core.
The term $\wtil{G}_{\rm av}$ is the pairing matrix element between active valence neutrons,
and $\wtil{G}_{\rm c-av}$ is the coupling between valence neutrons in \nuc{9}{Li} and 
active valence neutrons, which will be explained later in detail.
The matrix element of $\wtil{G}_{\rm pair}$ in \nuc{11}{Li} is given as:
\begin{eqnarray}
&&	\bra\Phi(C_{\al})\,\chi_\al(nn)|\wtil{G}_{\rm pair}|\Phi(C_{\al'})\,\chi_{\al'}(nn)\ket
	\nonumber
	\\
&=&	\bra\Phi(C_\al)|\wtil{G}_{\rm c}|\Phi(C_{\al'})\ket\! \cdot\! \bra\chi_\al(nn)|\chi_{\al'}(nn)\ket
+	\bra\Phi(C_\al)|\Phi(C_{\al'})\ket\! \cdot\! \bra\chi_\al(nn)|\wtil{G}_{\rm av}|\chi_{\al'}(nn)\ket
	\nonumber
	\\
&&	\hspace*{3cm}+~\bra\Phi(C_\al)\,\chi_\al(nn)|\wtil{G}_{\rm c-av}|\Phi(C_{\al'})\,\chi_{\al'}(nn)\ket
	\\
&=&	\left\{~
	\bra\Phi(C_\al)|\wtil{G}_{\rm c}|\Phi(C_{\al'})\ket 
~+~	\bra\chi_\al(nn)|\wtil{G}_{\rm av}|\chi_{\al'}(nn)\ket~
	\right\}\ \delta_{\al,\al'}
	\nonumber
	\\
&&	\hspace*{3cm}+~\bra\Phi(C_\al)\,\chi_\al(nn)|\wtil{G}_{\rm c-av}|\Phi(C_{\al'})\,\chi_{\al'}(nn)\ket,
	\label{eq:pair_abc}
\end{eqnarray}
where we omit the angular momentum coupling and antisymmetrization for simplicity.

Couplings between configurations (a)-(b) (active valence), and between (b)-(c) (core) are 
calculated via pairing coupling expressed in the first two terms of Eq.~(\ref{eq:pair_abc})
including $\wtil{G}_{\rm c}$ and $\wtil{G}_{\rm av}$.
The (a)-(c) (core-active valence) coupling is also evaluated by considering 
the exchange interaction between two neutrons in the \nuc{9}{Li} core and active valence neutrons,
explained in the previous paper\cite{Mu98}: 
Through the interaction, two valence neutrons of $(0p_{3/2})^2$ in the \nuc{9}{Li} core shown in (a) 
are exchanged with active valence neutrons excited to an $(nl_j)^2$-orbit shown in (c), 
and simultaneously the active valence neutrons of $(0p_{1/2})^2$-orbit in (a) are exchanged 
with two valence neutrons of $(0p_{1/2})^2$ in the \nuc{9}{Li} core shown in (c). 
This is expressed by the following matrix element:
\begin{eqnarray}
&&	\bra\Phi(C_0)\chi_{0p_{1/2}}(nn)|\wtil{G}_{\rm c-av}|\Phi(C_1)\chi_{nl_j}(nn)\ket
	\\
&=&	\bra\phi_{0p_{3/2}}(nn)\chi_{0p_{1/2}}(nn)|\wtil{G}_{\rm c-av}|\phi_{0p_{1/2}}(nn)\chi_{nl_j}(nn)\ket
	\\
&=&	\bra\phi_{0p_{3/2}}(nn)|\wtil{G}_{\rm c-av}|\chi_{nl_j}(nn)\ket\cdot
	\bra\chi_{0p_{1/2}}(nn)|\phi_{0p_{1/2}}(nn)\ket.
\end{eqnarray}
where $\phi_{nl_j}(nn)$ is the harmonic oscillator wave function of two valence neutrons in the \nuc{9}{Li} core.
The Minnesota interaction is used to calculate the exchange coupling of neutron-pairs.
For \nuc{10}{Li} such a neutron-pair exchange interaction does not appear.

We would like to mention about the difference between two kinds of the pairing correlations 
for the valence neutrons in the \nuc{9}{Li} core and for the active valence neutrons.
For the active valence neutrons, we adopt the hybrid-TV model in order to take into account 
the their pairing correlation microscopically.
On the other hand, for the \nuc{9}{Li} core, we truncate the configuration of the neutron paring excitation in 
\nuc{9}{Li} up to the $p$-shell, and employ the effective interaction to determine the pairing matrix element.
Therefore, treatments of two kinds of the pairing correlations are different in this study.
\section{Unified description of \nuc{10}{Li} and \nuc{11}{Li} in coupled-channel calculations}
\label{sec:result}

\subsection{Spectroscopy of \nuc{10}{Li}}\label{sec:10Li}
Before solving the extended three-body problem of \nuc{11}{Li},
we reinvestigate the spectroscopy of \nuc{10}{Li} and the \nuc{9}{Li}-$n$ interaction.\cite{Mu98,Ka93,Ka99}
The results of the positive parity states are shown in Table~\ref{tab:level_10Li_+},
where the active valence neutron dominantly occupies the $p$-wave.
We employ the complex scaling method\cite{ABC} to search the resonance poles.\cite{Ka93,Ka99,Ao97}
We fit the $1^+$ resonance with the experimental value; 0.42 MeV observed by Bohlen et al.,\cite{Bo93}
by adjusting the $\delta$ parameter to change
the strength of the second range in the \nuc{9}{Li}-$n$ interaction.

Since the strong pairing-blocking results in the strong \nuc{9}{Li}-$n$ interaction, 
we need a larger value of $\delta$ parameter
for the strong pairing correlation case (PC-S)
than that for the weak pairing correlation case (PC-W) 
as shown in Table~\ref{tab:level_10Li_+}.
It is found that we fairly reproduce the position of $2^+$ resonance,
and that the decay widths of $1^+$ states depend on the strength of pairing-blocking.
If pairing-blocking is strong, the decay widths becomes narrow and close to the experimental data.

\begin{table}[t]
\caption[$1^+$ and $2^+$ resonances of \nuc{10}{Li}.]{
$1^+$ and $2^+$ resonances of \nuc{10}{Li}.
Units are in MeV for resonance energies measured from the \nuc{9}{Li}-$n$ threshold 
and decay widths. Experimental data are taken from Ref.\protect\citen{Bo93}.
}       \label{tab:level_10Li_+}
\begin{center}
\begin{tabular}{c|c|c|c}
\hline\hline
    & PC-W & PC-S & Exp.\cite{Bo93}\\
\hline
$\delta$  & 0.0619 & 0.1573  & ---\\
\hline
$1^+~(E_r,\Gamma)$~&~~(0.42,0.22)~~&~~(0.42,0.16)~~&~~(0.42,0.15)~~\\
$2^+~(E_r,\Gamma)$~&~~(1.02,0.84)~~&~~(1.31,0.81)~~&~~(0.80,0.30)~~\\
\hline
\end{tabular}
\end{center}
\end{table}

\begin{table}[t]
\caption[Properties of the negative parity states in \nuc{10}{Li}]
{       Properties of negative parity states (virtual states) in \nuc{10}{Li}.
	Units are in MeV for energy measured from the \nuc{9}{Li}-$n$ threshold, 
	and fm for scattering length $a_s$.
}       \label{tab:s-wave_10Li}
\begin{center}
\begin{tabular}{c|c|c}
\hline\hline
  &~~~PC-W~~~&~~~PC-S~~~\\
\hline
$\delta$  & 0.0619 & 0.1573 \\
\hline
$E(1^-)$ & --- &  --- \\ 
$E(2^-)$ & --- & $-$0.38\\ 
\hline
$a_s(1^-)$ &  $+$1.83  & $+$0.2 \\
$a_s(2^-)$ &  $+$0.68  & $-$5.0 \\
\hline
\end{tabular}
\end{center}
\end{table}

In Table~\ref{tab:s-wave_10Li}, we list the properties of the negative parity states of \nuc{10}{Li}
having a $s$-wave component dominantly, in the cases of PC-W and PC-S.
The position of the virtual states having a negative imaginary momentum,
are calculated with so-called the Jost function method.\cite{Ma00,So97}
In the PC-W case, we cannot find any low energy virtual state, 
and the scattering length shows an almost zero value.
This means that there is no low energy $s$-wave state in \nuc{10}{Li} 
because the \nuc{9}{Li}-$n$ interaction for the $s$-wave is weak in the PC-W case.
On the other hand, in the PC-S case, the $2^-$ virtual state is obtained
near the threshold energy of \nuc{9}{Li}-$n$ system.
A negative value of the scattering length of this state also implies 
the presence of a low energy $s$-wave, 
which is consistent with the recent experimental observations.\cite{Zi95,Zi97,Th99}

\begin{figure}[b]
\begin{center}
\epsfxsize=11.5cm
\centerline{\epsfbox{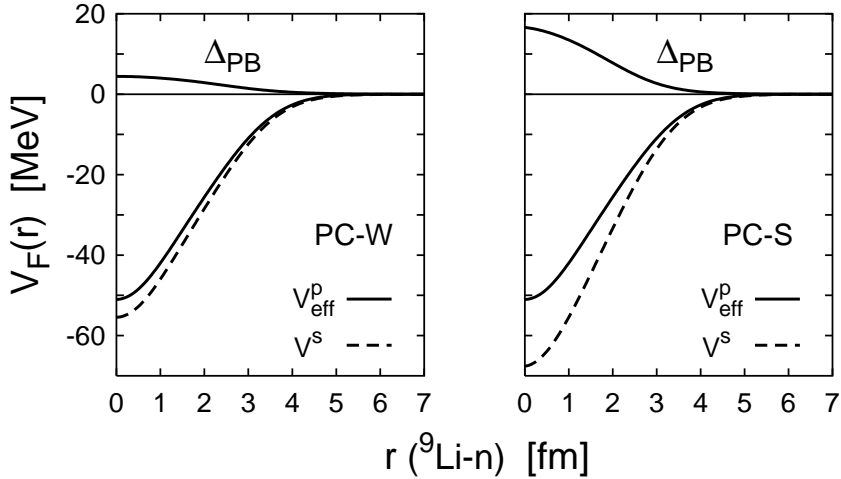}}
\caption[\nuc{9}{Li}-$n$ folding potential]{
\nuc{9}{Li}-$n$ folding potentials which are transformed into 
equivalent potentials defined in a single channel for only the $p$-wave.
The left panel is for the PC-W case and right one for the PC-S case.}
\label{fig:9Li-n_pot}
\end{center}
\end{figure}

In Fig.~\ref{fig:9Li-n_pot}, we draw the folding potentials for $s$- and $p$-waves.
For the $p$-wave, we make an equivalent effective potential $V_{\rm eff}^p$ to reproduce 
the position of the $1^+$ state as 0.42 MeV in the single channel calculation 
without the pairing correlation in \nuc{9}{Li}, namely, without the pairing-blocking effect.
This is done in the same manner as in Ref.~\citen{Ka99}. 
For the $s$-wave, the effective potential $V^s$ is given by the original folding potential.
We obtain the state-dependent interactions.
The $\Delta_{PB}$ values, which are the differences between the potentials of effective $p$-wave and of $s$-wave, 
reflect the strength of the pairing-blocking.
\begin{eqnarray}
	\Delta_{PB}
&\sim&  V_{\rm eff}^p-V^s.
\end{eqnarray}
From Fig.~\ref{fig:9Li-n_pot}, 
we find that the PC-S case leads to the deeper $s$-wave interaction.
Therefore, if we only use the idea of pairing-blocking coming from 
the pairing correlation in \nuc{9}{Li},
a stronger configuration mixing in \nuc{9}{Li} is favored
in order to reproduce the degeneracy of $s$- and $p$-waves in \nuc{10}{Li}.

\subsection{Ground state properties of \nuc{11}{Li}}\label{sec:11Li}
We perform the coupled-channel three-body calculation of \nuc{11}{Li}.
The important points in this calculation are whether the present model can solve
the underbinding problem, and describe the halo structure.
We also see how the pairing correlations work on the binding mechanism.
Results are shown in Table~\ref{tab:11Li}.
Energies of the \nuc{11}{Li} ground state measured from the three-body threshold 
are shown with switching on or off the exchange coupling explained in \S~\ref{sec:exchange}.
We can see that if there is no pairing correlation in \nuc{9}{Li} (fourth column),
namely, the calculation with the single configuration of the \nuc{9}{Li} core, 
the \nuc{11}{Li} ground state is not bound.
And by considering the pairing correlation in \nuc{9}{Li},
the \nuc{9}{Li}+$n$+$n$ system can be bound
due to the coupling to the pairing excited configuration of \nuc{9}{Li} in \nuc{11}{Li}. 
It is also found that the exchange coupling increases 
the probability of pairing excited configurations of \nuc{9}{Li} in \nuc{11}{Li} 
and also produces the energy gain by about 0.5 MeV, which
is close to the estimated value evaluated in Ref.~\citen{Mu98}. 
 
\nc{\lw}[1]     {\smash{\lower1.25ex\hbox{#1}}}
\begin{table}[b]
\caption[Properties of \nuc{11}{Li}]{
Properties of the \nuc{11}{Li} ground state.
Units are in MeV for the energy measured from the \nuc{9}{Li}+$n$+$n$ three-body threshold, 
and fm for r.m.s. radius.
The values in parenthesis are obtained by the calculation without exchange coupling. 
}\label{tab:11Li}
\begin{center}
\begin{tabular}{c|c|c|c|c}
\hline
\hline
          &~~\lw{PC-W}~~&~~\lw{PC-S}~~& No pairing~~ &~~\lw{Experiment}\\
          &             &             & correlation~~\\
\hline
$\delta$  & 0.0619 & 0.1573 & 0.0442  & ---\\
\hline
\lw{$E(3/2^-)$}    &  $-$0.50  &  $-$2.67  &       ---        & \lw{$-$0.31\cite{Au93}}\\ 
                   & ($-$0.11) & ($-$2.09) & ($0.70-i\,0.12$) & \\  
\hline
\lw{r.m.s. radius} &  2.69  &  2.49  & --- & $3.12\pm0.16$ fm\cite{Ta88b}\\ 
                   & (2.76) & (2.50) & --- & $3.53\pm0.06$ fm\cite{To97}\\
\hline
\lw{$(s_{1/2})^2$-probability} &  1.8\%    &  1.9\%  & --- & --- \\
                               & (2.7\%)   & (1.8\%) & --- & --- \\
\hline
\lw{$(p_{1/2})^2$-probability} &  94.4\%  &  93.9\%  & --- & --- \\
                               & (94.2\%) & (95.9\%) & --- & --- \\
\hline
$(a_1)^2$: probability of      &  1.5\%  &  2.0\%  & --- & ---\\ 
pairing excitation in \nuc{9}{Li}   & (0.9\%) & (0.5\%) & --- & ---\\
\hline
\end{tabular}
\end{center}
\end{table}

\begin{figure}[t]
\epsfxsize=12.0cm
\centerline{\epsfbox{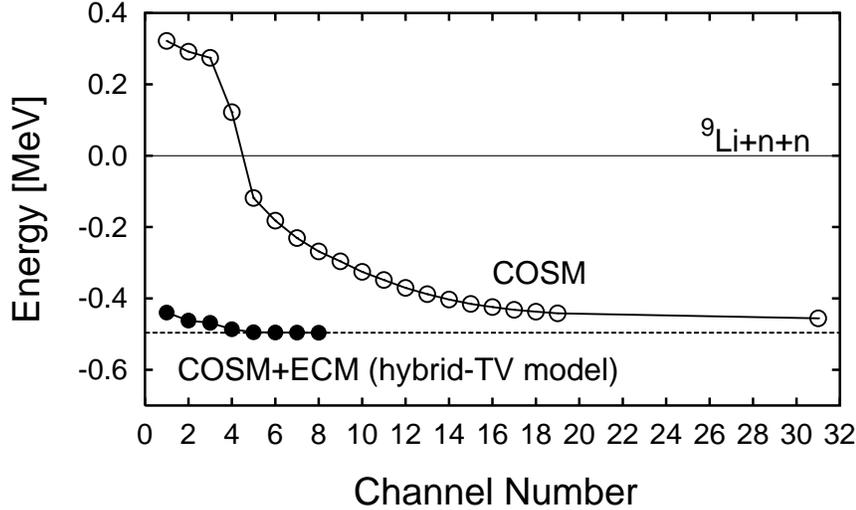}}
\caption[Convergence of energy of the \nuc{11}{Li} ground state]{
Convergence of the \nuc{11}{Li} ground state energy with respect to the channel numbers 
with only the COSM basis (open circles) and with the hybrid-TV one (solid circles).
Dotted line shows a converged energy ($-0.50$ MeV) in PC-W case.}
\label{fig:conv_ene}
\end{figure}

Before proceeding to the detail analysis of the \nuc{11}{Li} ground state, 
we discuss the role of the pairing correlation between active valence neutrons.
In Fig.~\ref{fig:conv_ene}, two kinds of the energy convergences of \nuc{11}{Li}
are shown with increasing the channel number of the $j^\pi=0^+$ pairing configuration for active valence neutrons.
One of them is the calculation with only the COSM basis and another is with the hybrid-TV ones including 
a T-type basis set. 
The parameter for the pairing correlation of PC-W is used including the exchange coupling.
In the calculation, we take the first channel as $(p_{1/2})^2$, and 
the order of added channels into the first one is 
$(s_{1/2})^2$,~$(p_{3/2})^2$,~$(d_{5/2})^2$,~$(d_{3/2})^2$, $\cdots$, $(l_j)^2$.
The maximum channel number is 31, where the orbital angular momentum and the spin of one active valence neutron are $(l,j)=(15,31/2)$.
We can see a rapid energy convergence in the hybrid-TV model rather than the COSM case.
This result indicates that the pairing correlation between active valence neutrons is important 
to reproduce the weak binding state of \nuc{11}{Li}.

Going back to the results in Table~\ref{tab:11Li},
PC-S gives larger binding energies than the experimental one (0.31 MeV).
This overbinding implies that the coupling to the pairing excited configurations of \nuc{9}{Li} is too strong. 
On the other hand, in the PC-W case, the calculated binding energies are closed to the experimental one. 
It is also found that the $(s_{1/2})^2$-probability is very small and the $p$-shell closed 
configuration is dominated in both cases of PC-W and PC-S, 
even if the weak binding energy is reproduced.
It is noticed that in the PC-S case, $s$- and $p$-waves are 
degenerated energetically in \nuc{10}{Li}. 
However the $(s_{1/2})^2$-probability is small.
The r.m.s. radius is still small in every cases of the calculations.
Even if we adjust the interaction to reproduce the experimental binding energy of \nuc{11}{Li}, 
this trend does not change. 

The probabilities of pairing excited configurations of \nuc{9}{Li} in the \nuc{11}{Li} ground state
are much smaller than that of the isolated \nuc{9}{Li} nucleus.
This indicates that in \nuc{11}{Li}, 
the energy gain from the potential energy between \nuc{9}{Li} and $p_{1/2}$-active valence neutrons 
is larger than that from the mixing of the pairing excited configurations in \nuc{9}{Li}.
However, this does not imply that the pairing correlation of \nuc{9}{Li} is unnecessary, 
even if the probability of pairing excitation in \nuc{9}{Li} is small. 
The pairing correlation of \nuc{9}{Li} is necessary to make 
the \nuc{11}{Li} ground state bound via the coupling to 
the wave function of \nuc{11}{Li} with $p_{3/2}$-closed configuration of \nuc{9}{Li}.

The reason why the $(s_{1/2})^2$-probability is very small, is explained 
by considering the expectation values of the Hamiltonian of $s$- and $p$-waves 
in \nuc{10}{Li} and \nuc{11}{Li}.
First we consider the case of \nuc{10}{Li} where the Hamiltonian consists of 
an internal part of the \nuc{9}{Li} cluster and the relative motion between \nuc{9}{Li} 
and an active valence neutron.
\begin{eqnarray}
     \bra H(\mbox{\nuc{10}{Li}},s)\ket
&=&  \bra H(\mbox{\nuc{9}{Li}},s)\ket + \bra H_{\rm rel}(s)\ket,
	\label{eq:expect_10Li_s}
     \\
     \bra H(\mbox{\nuc{10}{Li}},p)\ket
&=&  \bra H(\mbox{\nuc{9}{Li}},p)\ket + \bra H_{\rm rel}(p)\ket,
	\label{eq:expect_10Li_p}
\end{eqnarray}
where $H_{\rm rel}=T_{\rm rel} + V_{cn}$ and $H(\mbox{\nuc{9}{Li}})$ is given in Eq.~(\ref{eq:ham_9Li}),
and its expectation value $\bra H(\mbox{\nuc{9}{Li}},l)\ket$ includes the coupling effects 
with the $l$-orbital active valence neutron on the \nuc{9}{Li} core due to the Pauli-principle.
The relations of each term between $s$- and $p$-waves are given as:
\begin{eqnarray}
     \bra H(\mbox{\nuc{9}{Li}},p)\ket
~&>&~\bra H(\mbox{\nuc{9}{Li}},s)\ket,
	\qquad
     \bra H_{\rm rel}(p)\ket
~<~  \bra H_{\rm rel}(s)\ket.
	\label{eq:expect_10Li}
\end{eqnarray}
First relation comes from the fact that appearance of the $p_{1/2}$-active valence neutron changes
the configuration mixing of \nuc{9}{Li} from that of the isolated \nuc{9}{Li}
due to the Pauli-principle, and then the \nuc{9}{Li} core loses its energy.
On the other hand, for the $s$-wave active valence neutron, mixing in \nuc{9}{Li} core 
is not disturbed in \nuc{10}{Li}, and the energy loss does not occur.
In the second relation, the relative motion of the $p$-wave is energetically gained 
rather than that of the $s$-wave because the $s$-wave neutron confined in the interaction region 
has a larger kinetic energy than that of the $p$-wave. 
In \nuc{10}{Li} each term of the expectation values are canceled mutually 
in both cases of $s$- and $p$-waves.
If the pairing-blocking effect is sufficiently strong 
to eliminate the pairing excited configuration of \nuc{9}{Li}, 
the gap of the expectation values of the \nuc{9}{Li} Hamiltonian 
in the first relation in inequality (\ref{eq:expect_10Li})
becomes large, 
and total expectation values of \nuc{10}{Li} Hamiltonian for the $s$- and the $p$-waves can be degenerated.

Next we consider the case of \nuc{11}{Li}.
Expectation values of the Hamiltonian of the three-body system are given as:
\begin{eqnarray}
     \bra H(\mbox{\nuc{11}{Li}},s)\ket
&=&  \bra H(\mbox{\nuc{9}{Li}},s)\ket + 2\, \bra H_{\rm rel}(s)\ket + \bra H_{nn}(s)\ket
     \\
&=&  \bra H(\mbox{\nuc{10}{Li}},s)\ket + \bra H_{\rm rel}(s)\ket + \bra H_{nn}(s)\ket,
	\label{eq:expect_11Li_s}
    \\
     \bra H(\mbox{\nuc{11}{Li}},p)\ket
&=&  \bra H(\mbox{\nuc{9}{Li}},p)\ket + 2\, \bra H_{\rm rel}(p)\ket + \bra H_{nn}(p)\ket
     \\
&=&  \bra H(\mbox{\nuc{10}{Li}},p)\ket + \bra H_{\rm rel}(p)\ket + \bra H_{nn}(p)\ket,
	\label{eq:expect_11Li_p}
\end{eqnarray}
where $H_{nn}= V_{nn} + \frac{\vc{p}_1\cdot\vc{p}_2}{9m}$ including the cross term of kinetic energy,\cite{To90a}
and Eq.~(\ref{eq:expect_11Li_s}) and (\ref{eq:expect_11Li_p}) are derived using 
Eq.~(\ref{eq:expect_10Li_s})  and (\ref{eq:expect_10Li_p}).
We checked that the pairing interaction $V_{nn}$ between two active valence neutrons 
is more attractive for the $p$-wave than for the $s$-wave. 
The difference between $\bra V_{nn}(s)\ket$ and $\bra V_{nn}(p)\ket$ is about 0.5 MeV in this calculation.
Expectation values of the cross term of the kinetic energy is zero due to the spin condition
(there is an off-diagonal coupling of $s^2$-$p^2$, which is small).
The first terms in Eqs.~(\ref{eq:expect_11Li_s}) and (\ref{eq:expect_11Li_p})
can be degenerated as was explained above.
Therefore,  using the second relation in inequality~(\ref{eq:expect_10Li}),
we can easily notice the following relation of the expectation values of 
the Hamiltonian of \nuc{11}{Li}:
\begin{eqnarray}
     \bra H(\mbox{\nuc{11}{Li}},p)\ket
~&<&~\bra H(\mbox{\nuc{11}{Li}},s)\ket.
     \label{eq:expect_11Li}
\end{eqnarray}
It is found that the $p$-shell dominant configuration is favored in the \nuc{11}{Li} ground state.

In the case of the KYI parameter in which the $s$- and the $p$-waves are well degenerated in \nuc{10}{Li},
the maximum gap of the first relation in Inequality (\ref{eq:expect_10Li})
is estimated as 3.25 MeV from Table~\ref{tab:pairing-core}.
If we assume the same expectation values of the \nuc{10}{Li} Hamiltonian for $s$- and $p$-waves, 
the energy gap between $(s_{1/2})^2$- and $(p_{1/2})^2$-components 
of \nuc{11}{Li} is naively estimated to be around 3-4 MeV. 

From these relations shown in inequalities~(\ref{eq:expect_10Li}) and (\ref{eq:expect_11Li}),
we can find that the pairing correlation in \nuc{9}{Li} 
provides the energetical degeneracy of $s$- and $p$-wave active valence neutron in \nuc{10}{Li},
but the energies of $(p_{1/2})^2$ and $(s_{1/2})^2$-orbital neutrons in \nuc{11}{Li}
are not degenerate.
The $(s_{1/2})^2$-probability in the \nuc{11}{Li} ground state shows no enhancement. 
Indeed, underbinding problem seems to be solved in the present model, 
but, the wave function of the \nuc{11}{Li} ground state is not sufficient 
to reproduce the halo structure.

\begin{figure}[b]
\epsfxsize=12.5cm
\centerline{\epsfbox{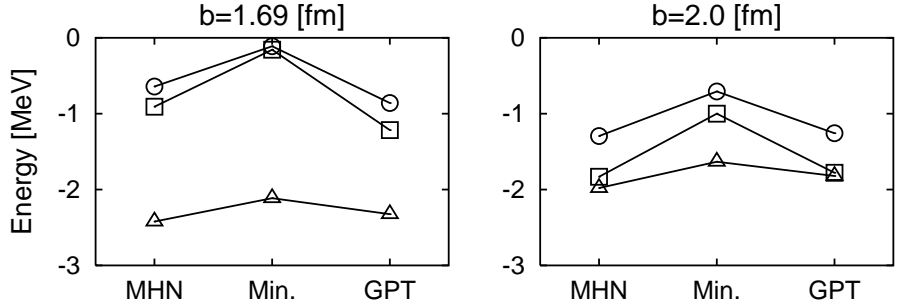}}
\caption[Paring matrix elements between 0p-1s waves]{
Paring matrix elements between 0p-1s waves of harmonic oscillator basis function 
with two length parameters b=1.69 (left) and 2.0 (right) fm.
Circles, squares, and triangles indicate 
$(-4)\times G_{0p_{1/2},1s_{1/2}}$,
$(-4)\times G_{0p_{3/2},1s_{1/2}}$,
$G_{1s_{1/2},1s_{1/2}}$, respectively.}
\label{fig:s-p_matrix}
\end{figure}

\subsection{S-P pairing coupling in \nuc{11}{Li}}\label{sec:s-p}

In the previous subsection, we discuss the relations of 
the diagonal matrix elements of the Hamiltonian for $s$- and $p$-waves 
in \nuc{10}{Li} and \nuc{11}{Li}, respectively.
Here, we also examine the coupling matrix element between these partial waves in \nuc{11}{Li}, 
where only the interaction between active valence neutrons can contribute in the present three-body model.
Coupling between $s$- and $p$-waves is important to discuss the mixing of $s$-wave to form the halo structure 
in the \nuc{11}{Li} ground state.
In Fig.~\ref{fig:s-p_matrix}, we show the pairing matrix element of $0p$-$1s$ waves 
in the harmonic oscillator wave function with two length parameters. One is 1.69 fm, 
the same as that of \nuc{9}{Li} and another is 2.0 fm corresponding to a large value of the halo distribution.
In addition to the MHN and the Minnesota interactions, the GPT one including tensor and LS parts is also compared.
We already check that the GPT interaction for active valence neutrons gives the similar results for \nuc{11}{Li}, 
such as binding energy and $(s_{1/2})^2$-probability as those of the Minnesota one.
From Fig.~\ref{fig:s-p_matrix}, it is found that the Minnesota interaction has a tendency to give 
the small coupling matrix elements of the $0p$-$1s$ part than other interactions,
in particular, when length parameter is small. 
On the other hand, the diagonal matrix elements of $1s$-waves show similar values among these interactions.
This result implies that there is an ambiguity to estimate the 
coupling matrix element of $\bra (0p)^2|V_{nn}|(1s)^2\ket$.

\begin{figure}[t]
\epsfysize=6.2cm
\centerline{\epsfbox{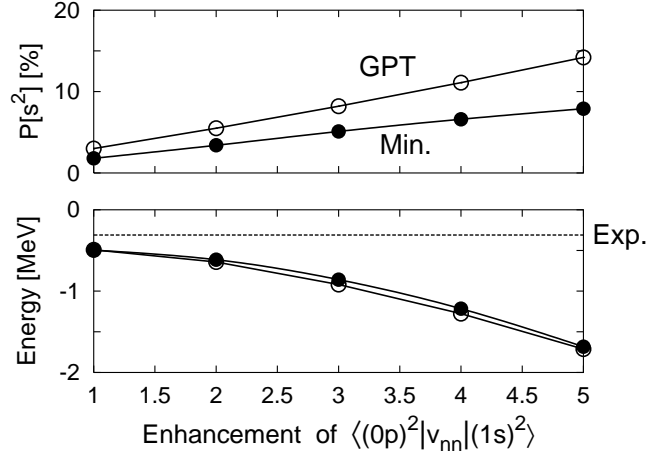}}
\caption[Dependence of energy and matter radius of \nuc{11}{Li}]{
Dependences of energy and matter radius of \nuc{11}{Li} 
on the enhancement of $\bra (0p)^2|v_{nn}|(1s)^2\ket$ with Minnesota interaction (solid circles)
and GPT one (open circles).}
\label{fig:s-p}
\end{figure}

In Fig.~\ref{fig:s-p}, we 
see the dependences of binding energy and $(s_{1/2})^2$-probability of the \nuc{11}{Li} ground state
on the pairing matrix element of this coupling term.
Minnesota and GPT interactions are used for the active valence neutrons.
GPT interaction gives the larger $(s_{1/2})^2$-probability due to the large coupling matrix 
element of $s^2$-$p^2$ than that of Minnesota.
We find the fact that if we strengthen the coupling to five times,
although the $(s_{1/2})^2$-probability becomes larger around 10\% to 15\%, 
an overbinding problem appears with two interactions.

Here, we shortly summarize the results of \nuc{10}{Li} and \nuc{11}{Li}.
The pairing correlation in \nuc{9}{Li} nicely works to reproduce the low-energy properties of \nuc{10}{Li} 
as a result of the pairing-blocking,
and becomes a nice reason to suggest the state-dependent \nuc{9}{Li}-$n$ interaction.
In \nuc{11}{Li}, however, we obtain a three-body bound state dominated by the $p$-shell closed configuration.
This result indicates the pairing correlation works differently in 
odd- and even-neutron nuclei of \nuc{10}{Li} and \nuc{11}{Li}, 
namely, the couplings between active valence neutrons and the \nuc{9}{Li} core
are different in \nuc{10}{Li} and \nuc{11}{Li}.

In order to improve the configuration of the \nuc{11}{Li} ground state,
we must increases the coupling of the pairing matrix element $\bra (0p)^2|v_{nn}|(1s)^2\ket$. 
But, overbinding problem appears by increasing the coupling,
and this problem is difficult to solve at the present stage.
Then we may conclude that the present three-body model 
with taking into account the pairing correlation in \nuc{9}{Li}
is not enough to reproduce the low-energy properties of \nuc{10}{Li} and \nuc{11}{Li} simultaneously.

However, this conclusion does not mean that the present approach 
is wrong. Our model well reproduces the properties of \nuc{10}{Li}, and
furthermore, there is room to discuss 
the improvements of the present model in addition to the pairing correlation,
as will be discussed in the next section.
\section{Discussion on the tail behavior of the \nuc{9}{Li}-$\mib{n}$ interaction}\label{sec:discuss}

We showed that our extended three-body model of \nuc{11}{Li} cannot reproduce 
a large r.m.s. radius and a large amplitude of the $(s_{1/2})^2$ component,
consistently with the spectroscopy of \nuc{10}{Li}.
For the properties of $s$-waves,
it is noticed that the behaviour of the $s$-wave state near the threshold energy is 
sensitive to the tail part of the interaction due to the 
spatially extended distribution of the wave function without the centrifugal barrier.
In the present model, we adopt the folding potential as the \nuc{9}{Li}-$n$ interaction 
in a Gaussian form which rapidly falls off with a large distance.
Also we assume the harmonic oscillator wave function of the \nuc{9}{Li} core 
and use the effective $NN$ interaction of Gaussian form.
It is worthwhile to investigate the effect of the tail part of the \nuc{9}{Li}-$n$ interaction
on the $s$-wave states in \nuc{10}{Li} and the mixing of $s$-waves in the \nuc{11}{Li} ground state .

\begin{figure}[b]
\epsfysize=6.0cm
\centerline{\epsfbox{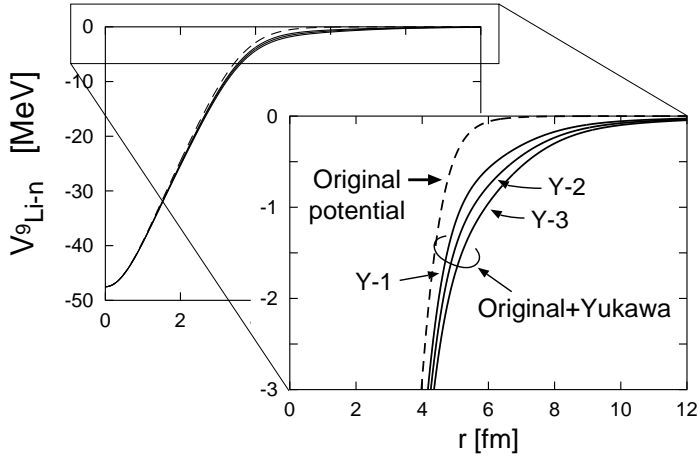}}
\caption[]{
\nuc{9}{Li}-$n$ interaction including tail effect.}
\label{fig:tail}
\end{figure}

For the expression of the tail part,
we add a phenomenological tail potential to the original folding one for \nuc{9}{Li}-$n$.
We employ the Yukawa-type form with length parameter $b_t$ as 2.4 fm, 
calculated from the neutron separation energy of \nuc{9}{Li}, about 4.1 MeV. 
\begin{eqnarray}
	V_t
&=&	v_t\ \frac{e^{-r/b_t}}{r} \times W(r),
\end{eqnarray}
where $W(r)$ is a weight function which is chosen to be zero inside the \nuc{9}{Li} nucleus 
( less than around 3 fm),
and unity in the outside region beyond the surface of the \nuc{9}{Li} core.
The form of the \nuc{9}{Li}-$n$ interaction with tail part is drawn in Fig.~\ref{fig:tail}.
Here, we chose three types of the strengths; Y-1 ($v_t$=$-$40 MeV), Y-2 ($-$55 MeV) and Y-3 ($-$65 MeV). 
Although the magnitude of tail interaction is very small compared with 
the original folding potential, it is shown that its effect on the structure of \nuc{11}{Li} is large.
The $\delta$ parameter in the folding potential is determined to reproduce the experimental
$1^-$ state in \nuc{10}{Li}, in the same manner as that of the previous section.

We also introduce a parameter $\alpha_{s\mbox{-}p}$ to change the $(0p)^2$-$(1s)^2$ pairing coupling as:
\begin{eqnarray}
	\bra (0p)^2|v_{nn}|(1s)^2\ket
~&\to&~  \alpha_{s\mbox{-}p}\cdot \bra (0p)^2|v_{nn}|(1s)^2\ket,
\end{eqnarray}
because, we mentioned that there is an ambiguity to estimate this coupling matrix elements in the $NN$ interaction,
and it might be expected that the pairing correlation can be changed in \nuc{11}{Li} due to the 
coupling of two kinds of the neutron pairings of \nuc{9}{Li} core part and active valence part. 
The value of $\alpha_{s\mbox{-}p}$ is determined to reproduce the experimental binding energy of \nuc{11}{Li}.

In Table.~\ref{tab:MF_10Li} and \ref{tab:MF_11Li}, 
we list the results of \nuc{10}{Li} and \nuc{11}{Li},
where PC-W is used as the pairing correlation in \nuc{9}{Li}, 
and the MHN interaction for the \nuc{9}{Li}-$n$ folding potential.
For \nuc{10}{Li}, it is found that 
all $s$-states ($1^-$,~$2^-$) are obtained as virtual ones having small energies 
and show the large negative scattering lengths.
The difference of Y-1,Y-2, and Y-3 is well reflected in the values of scattering lengths.
These results means that the tail interaction gives a large effect to improve the properties of $s$-wave. 
On the other hand, the decay width of the $p$-states become larger than those in Table~\ref{tab:level_10Li_+}.
It is noticed that due to the introduction of the tail interaction,
we do not need the strong pairing-blocking, namely, a large $\delta$ value in the 
\nuc{9}{Li}-$n$ folding potential.

\begin{table}[t]
\caption[]{
        Results of \nuc{10}{Li} with the pairing correlation of \nuc{9}{Li} and the tail effect. 
        Unit in MeV for energy and fm for scattering length $a_s$
        the negative parity states of \nuc{10}{Li} are virtual one.
}       \label{tab:MF_10Li}
\begin{center}
\begin{tabular}{c|ccc}
\hline
\hline
parameter set of the tail interaction  & Y-1 &  Y-2 & Y-3 \\
\hline
$\delta$ & 0.0290 & 0.0132 & 0.0009 \\
\hline
$1^+~(E_r,\Gamma)~$ &~(0.42,0.33)~&~(0.42,0.38)~&~(0.42,0.43)~\\
$2^+~(E_r,\Gamma)~$ &(0.79,1.2) &(0.64,1.2)  &(0.54,1.1)\\
\hline
$E(1^-) $ & $-$0.12 & $-$0.08 & $-$0.07 \\
$E(2^-) $ & $-$0.09 & $-$0.05 & $-$0.04 \\
\hline
$a_s(1^-)$ & $-$2.9 & $-$5.5  & $-$7.5  \\
$a_s(2^-)$ & $-$6.3 & $-$10.8 & $-$14.8 \\
\hline
\end{tabular}
\end{center}
\end{table}
\begin{table}[t]
\caption[]{
        Results of \nuc{11}{Li} with the pairing correlation in \nuc{9}{Li} and the tail effect. 
        Unit in MeV for energy and fm for r.m.s. radius.
}       \label{tab:MF_11Li}
\begin{center}
\begin{tabular}{c|ccc}
\hline
\hline
parameter set of the tail interaction & Y-1 & Y-2 & Y-3\\
\hline
$E(3/2^-)$       & $-$0.31 & $-$0.31 & $-$0.31 \\
\hline
r.m.s. radius  & 2.99 & 3.24 & 3.47 \\
\hline
$(s_{1/2})^2$-probability   & 9.1\% &  17.4\% & 25.1\% \\
\hline
$(p_{1/2})^2$-probability &~~85.2\%~~&~~75.3\%~~&~~66.1\%~~\\
\hline
$(a_1)^2$ : probability of         & \lw{3.2\%} & \lw{4.8\%} & \lw{6.2\%} \\
pairing excitation in \nuc{9}{Li}  & & & \\
\hline
$\alpha_{s\mbox{-}p}$ & 2.0 & 2.5 & 2.8 \\
\hline
\end{tabular}
\end{center}
\end{table}

For \nuc{11}{Li},
the $(s_{1/2})^2$-probability in the ground state 
is increased enough to reproduce the experimental matter radius, 
$3.12\pm0.16$ fm\cite{Ta88b} and $3.53\pm0.06$ fm.\cite{To97}
In our model, even a smaller $(s_{1/2})^2$-probability easily enhances the matter radius of \nuc{11}{Li}.
This is due to the tail effect in the \nuc{9}{Li}-$n$ interaction and wave function of the $s$-wave 
is easily spatially extended.
It should be mentioned that the tail interaction plays a important role in \nuc{11}{Li}
to lower the energy of $(s_{1/2})^2$-component with respect to that of $(p_{1/2})^2$-component.
The Y-3 case gives the lowest energy of $(s_{1/2})^2$-component. 
As a result, $(s_{1/2})^2$- and $(p_{1/2})^2$-components are easy to couple in the \nuc{11}{Li} ground state
by adjusting the $\alpha_{s\mbox{-}p}$ parameter.
It is found that the adjusted values of $\alpha_{s\mbox{-}p}$ become larger than two.
This might indicate the enhancement of the pairing correlation 
due to the coupling of neutrons between active valence and core parts.

It is also found that the probabilities of pairing excited configurations of \nuc{9}{Li} in the \nuc{11}{Li} ground state
are larger than those of the previous results listed in Table~\ref{tab:11Li} by few percents.
This means that the pairing-blocking effect due to the active valence neutrons is relaxed in \nuc{11}{Li},
because the $(p_{1/2})^2$-probability of active valence neutrons decreases.
\section{Summary and conclusion
}\label{sec:summary}
We analyzed the structure of the \nuc{11}{Li} ground state 
and the low-energy structure of \nuc{10}{Li}
with the extended three-body model of \nuc{9}{Li}+$n$+$n$, 
where we describe the \nuc{9}{Li} cluster as the multi-configuration 
with paying attention to the pairing correlation of neutrons.
In \nuc{10}{Li}, the pairing correlation of the \nuc{9}{Li} core produces 
the so-called pairing-blocking effect due to the presence of the active valence neutron, 
which successfully works to reproduce the spectroscopy of \nuc{10}{Li}, 
in particular, the degeneracy of $s$- and $p$-wave states around the threshold energy.
The state-dependency of the \nuc{9}{Li}-$n$ interaction can be effectively derived 
within the description of \nuc{10}{Li} in our model.

In \nuc{11}{Li}, we met the underbinding problem in the previous study.
By employing the idea of the pairing-blocking, 
it is naively expected that the lacking of binding energy is recovered and the halo structure appears, 
since $s$- and $p$-waves are degenerate in \nuc{10}{Li}.
The result of the present coupled-channel three-body calculation of \nuc{11}{Li}
shows that the underbinding problem is solved.
The obtained binding energy depends on the strength of the pairing correlation in \nuc{9}{Li}.
If we employ the strong pairing mixing which is good for the description of \nuc{10}{Li},
the \nuc{11}{Li} ground states is overbound by 2 MeV, which is close to an estimated 
value (1.5 MeV\cite{Ao02} + 0.5 MeV\cite{Mu98}) within a conventional \nuc{9}{Li}+$n$+$n$ model.
Furthermore, the $s^2$-probability is too small 
in any case of the strength of pairing correlation in \nuc{9}{Li}.
This different result from \nuc{10}{Li} is 
because of the additional degrees of freedom of one more active neutron in \nuc{11}{Li}. 
Due to the additional active valence neutron, 
the pairing-blocking effect on the pairing correlation in \nuc{9}{Li} gives way to 
the attractive \nuc{9}{Li}-$n$ interaction for the $p_{1/2}$-orbit in \nuc{11}{Li}.
From this study, we can conclude that the effect of the pairing correlation 
are different in \nuc{10}{Li} and in \nuc{11}{Li}, namely in the odd- and even-neutron systems.
This indicates that one must be careful to use commonly the state-dependent \nuc{9}{Li}-$n$ potential 
in the calculations of \nuc{10}{Li} and \nuc{11}{Li}, 
even if such an approach is successful to explain the properties of these two nuclei simultaneously.\cite{Ao02}

We further discussed the effect of the tail of the \nuc{9}{Li} core.
The results using the \nuc{9}{Li}-$n$ interaction including the tail effect show that 
the mixing of $s$-waves is sufficiently improved and a large r.m.s. radius of \nuc{11}{Li} 
is well reproduced.

For the coupling between the \nuc{9}{Li} core and active valence neutrons including pairing interaction,
it is noticed that the antisymmetrization between active valence neutrons 
and neutrons in \nuc{9}{Li} is taken into account within the framework of OCM in this study,
even if we introduce the coupling due to the exchange of two neutron pairs 
explained in \S~\ref{sec:exchange}.
The dynamical effects such as the structure change of \nuc{9}{Li} core in \nuc{11}{Li} 
due to the Pauli-principle are dropped out.
It is expected to investigate such a dynamical effect which is related to the self-consistency of neutrons in \nuc{11}{Li}.
This is not considered in the present three-body model and is also beyond the task of this study.

For the experimental information of $p$-wave resonances of \nuc{10}{Li},
several experimental groups reported the different values from 
the one we used, 0.42 MeV for $1^+$ state.\cite{Yo94,Bo99}
Although the spin of these states is not fixed yet,
0.54 MeV for $2^+$ and 0.24 MeV for $1^+$ are suggested.
If we choose these values in the \nuc{9}{Li}+$n$ model, we can obtain a larger binding energy of \nuc{11}{Li}.  
However, the essential points about the effect of the pairing correlation 
for \nuc{10}{Li} and \nuc{11}{Li} obtained in this study, do not change. 

Although we discussed the pairing effects, there are other possibilities of 
the mechanism to degenerate the $s$- and $p$-waves in neutron-rich nuclei except for the pairing-blocking. 
Recently, Otsuka {\em et al.} proposed the importance of the spin-isospin part of the $NN$ interaction 
in the inversion problem of neutron-rich $N=7$ isotone based on the shell model.\cite{Ot01}
And the degrees of freedom of the deformation of the \nuc{9}{Li} core might also 
affect on the motion of the active valence neutron in \nuc{10}{Li}, although it is expected to be small.
Indeed these contributions including the pairing-blocking 
would combine with each other to produce the inversion phenomena in a real situation, 
we emphasize the roles of the pairing-blocking in this study.

As a next step, we will analyze the Coulomb breakup reaction of \nuc{11}{Li} 
to see the structures of the three-body unbound states of \nuc{11}{Li} 
including any resonances and continuum states.
It is very interesting to investigate the possibility of a soft-dipole resonance
because the low-lying $s$-states in \nuc{10}{Li}
can be expected to make the low-lying excited states in \nuc{11}{Li}. 
We will adopt the complex scaling method to describe those unbound states, 
which we succeeded in the application of the \nuc{6}{He} Coulomb breakup 
into \nuc{4}{He}+$n$+$n$.\cite{My01}

\section*{Acknowledgements}
The author would like to acknowledge valuable discussions with 
Professor G.~F.~Filippov, Professor S. Okabe, and Professor A. Ohnishi. 
The author also would like to thank the members of the nuclear theory 
group in Hokkaido University for their kind interest and discussions. 
One of the authors (T.M.) thanks Dr. H. Masui and Ms. Kurokawa 
for help to use the Jost function method.
This work is supported by the Grant-in-Aid for Scientific Research (No.~12640246)
of the Ministry Education, Science and Culture, Japan.
One of the authors (T.M.) thanks to the Japan society for the promotion of science
for support.
This work was performed as a part of the "Research Project for Study of
Unstable Nuclei from Nuclear Cluster Aspects (SUNNCA)" sponsored by 
Institute of Physical and Chemical Research (RIKEN).

\end{document}